\documentclass[reprint,pra,aps,
superscriptaddress,
amsmath,amssymb,
]{revtex4-1}

\usepackage{color}
\usepackage{graphicx}
\usepackage{subcaption}
\usepackage{dcolumn}% Align table columns on decimal pt
\usepackage{bm}% bold math
\usepackage{tabularx} %tables
\usepackage{hyperref}% add hypertext capabilities
\usepackage{tablefootnote}
\newcommand{\rg}{r_{\rm g}}
\newcommand{\wg}{\omega_{\rm g}}

\begin{document}

%---------------------------------------------------
%\newcommand{\kt}{\sqrt{k\tau}}
%\newcommand{\FWHM}{\sigma_{\text{FWHM}}}
%\newcommand{\Psim}{\left(1-\Psi\right)}
%\newcommand{\Psip}{\left(1+\Psi\right)}
%---------------------------------------------------

\bibliographystyle{unsrt}
%\preprint{APS/123-QED}

\title{Analytical estimates of proton acceleration in laser-produced turbulent plasmas}
%\thanks{A footnote to the article title}

\author{Konstantin A. Beyer}
\email{Correspondence to konstantin.beyer@physics.ox.ac.uk}
\affiliation{Department of Physics, University of Oxford, Parks Road, Oxford OX1 3PU, UK}
\author{Brian Reville}
\affiliation{School of Mathematics and Physics, Queens University Belfast, Belfast BT7 1NN, UK}
\author{Archie F. A. Bott}
\affiliation{Department of Physics, University of Oxford, Parks Road, Oxford OX1 3PU, UK}
\author{Hye-Sook Park}
\affiliation{Lawrence Livermore National Laboratory, P.O. Box 808, Livermore, California 94551, USA}
%\author{John Foster}
%\affiliation{AWE, Aldermaston, Reading, West Berkshire RG7 4PR, UK}
\author{Subir Sarkar}
\affiliation{Department of Physics, University of Oxford, Parks Road, Oxford OX1 3PU, UK}
\affiliation{Niels Bohr Institute, Blegdamsvej 17, 2100 Copenhagen, Denmark}
\author{Gianluca Gregori}
\affiliation{Department of Physics, University of Oxford, Parks Road, Oxford OX1 3PU, UK}
\date{\today}

\begin{abstract}
With the advent of high power lasers, new opportunities have opened up for simulating  astrophysical processes in the laboratory. We show that 2nd-order Fermi acceleration can be directly investigated at the National Ignition Facility, Livermore. This requires measuring the momentum-space diffusion of 3 MeV protons produced within a turbulent plasma generated by a laser. Treating  Fermi acceleration as a biased diffusion process, we show analytically that a measurable broadening of the initial proton distribution is then expected for particles exiting the plasma.
\end{abstract}

\maketitle

\section{\label{sec:level1} Introduction}
Turbulent magnetic fields are ubiquitous in the universe and their role in determining energetic particle transport is key to understanding the confinement and acceleration of high-energy cosmic rays \citep{TurbForStochAcc,ImpPlasmaProb,VoidMagnetTurb,TurbAcceleration,gregori2015the}. As particles traverse a turbulent, magnetised plasma, they undergo a random walk in both physical and momentum space. The latter process is referred to as 2nd-order Fermi acceleration, being a generalization of the mechanism proposed by Fermi \citep{FermiAcceleration}. Fermi observed that repeated elastic scattering of fast particles off slow moving `magnetised clouds', when averaged over a random distribution of cloud velocities produce a net gain in energy. The rate of energy gain is slightly higher than the rate of energy loss because head-on collisions are more probable than overtaking ones, the net gain being proportional to $(u/v)^2$ where $u$ is the  mean fluid velocity and $v (\gg u)$ the particle velocity. Subsequently the focus has shifted from discrete interactions with magnetised clouds to continuous scattering in magneto-hydrodynamic (MHD) turbulence, but the underlying principle remains the same. Fermi noted shortly afterwards that converging flows such as might exist between Galactic spiral arms result in a faster 1st-order process where the energy gain is proportional to $u/v$ \citep{Fermi54}. Indeed 1st-order Fermi acceleration in the converging fluid flow at astrophysical shock waves is currently the preferred model for the origin of cosmic rays \citep{Bell1,1978ApJ...221L..29B,ParticleAccelerationTheory}. However the 2nd-order mechanism can be more efficient for accelerating non-relativistic thermal background plasma particles and under certain conditions can also preferentially accelerate electrons relative to protons as is required to explain many astrophysical sources \citep{FermiReview}. In reality there may be a hybrid mechanism, e.g. initial 2nd-order acceleration of background plasma particles by turbulence, followed by a second stage of 1st-order acceleration by a shock wave.

The 2nd-order Fermi process is quite general, requiring only turbulent magnetised fluid motions and injection of particles with energy above that of the background thermal plasma. Relevant environments are common in the universe and stochastic acceleration is believed to be responsible for phenomena as diverse as e.g. radio emission from young supernova remnants entering the Sedov-Taylor phase \citep{SecOrdFermSolAltern}, the ejection of mass from the Solar corona \citep{1985srph.book..333N}, the acceleration of particles in the jets of active galactic nuclei and in their giant radio lobes \citep{Tramacere07,OSullivan09,Hardcastle09} and $\gamma$-ray emission from the  Fermi bubbles \citep{2011PhRvL.107i1101M}.

The necessary conditions may be accessible in laboratory experiments \citep{gregori2015the} thus providing a platform to explore particle acceleration in a controlled setting and isolate effects of relevance to astrophysical models. 
We explore here the possibility of validating the physics of 2nd-order Fermi acceleration using existing experimental set-ups (see supplementary material to Ref.\citep{tzeferacos2018laboratory}. 
In \S~\ref{sec:level2} we introduce the proposed set-up and place it in the context of previous experiments. The governing equations for the momentum space diffusion process are stated in \S~\ref{sec:level3} and the relevant Fokker-Planck coefficients of the diffusion process are estimated. We discuss the relevant time scales to justify the diffusion approach adopted. An analytic solution for the diffusion equation is investigated in \S~\ref{sec:level4}. Finally in \S~\ref{sec:level5}, laser experiments at the National Ignition Facility (NIF), Livermore, USA~\citep{moses} are discussed. We conclude that the effects of stochastic Fermi acceleration are measurable in the laboratory.

%-------------------------------------------------------- 
\section{\label{sec:level2} Experimental set-up}
Experiments with high power lasers have achieved conditions where strong magnetised turbulence can be sustained over large spatial scales and thus provide insights on the origin and amplification of magnetic fields in the intergalactic medium \citep{meinecke2014turbulent,meinecke2015developed, gregori2015the}. \citep{tzeferacos2018laboratory} describe how a high power laser was used to generate two counter streaming plasma flows from direct ablation of CH (plastic) foils. Each flow was guided through a grid of $300~\mu$m holes with $300~\mu$m spacing between holes. The grids were spatially shifted to increase turbulent motions in the plasma and enhance the turbulent dynamo processes responsible for amplification of magnetic seed fields. This produced turbulent structures with an outer scale of $\sim 600~\mu$m in the colliding region. These results were obtained using multi-kJ laser systems.
At the NIF, a MJ of laser energy is available so more extreme conditions are to be expected as shown in Table~I.
\begin{table}
\begin{center}
\def~{\hphantom{0}}
  \begin{tabular}{lccc}
  RMS magnetic field & $B\sim 1.2\,$MG \\
  Mean turbulent velocity & $u=6\times 10^7\,$cm/s \\
  Scale of the turbulence cells & $\ell\sim 0.06\,$cm \\
  Plasma size & $L=0.4\,$cm \\
  Initial proton momentum & $p_0=0.002\,m_{\rm p} c$ \\
  Temperature & $T=700\,$eV \\
  Electron density & $n=7\times 10^{20} \,\text{cm}^{-3}$\\
  Density relation & $\Delta n / n\sim O(1)$ \\
  Plasma beta \tablefootnote{The dependence on B is made explicit here, as we consider different values throughout the paper. All other parameters are as stated in Table~I. $\beta=4\times 10^{-11} nT/B^2$ in Gaussian cgs.} & $\beta=13.7\,\left(1.2\,\text{MG} / B\right)^2$\\ %=\frac{8\pi n T}{B^2}
  Alfv\'enic Mach number & $M_{\rm a}=u/v_{\rm a}=6$\\
  Reynolds number & $Re=1200$\\
  Magnetic Reynolds number & $R_{\rm m}=25000$\\
  \end{tabular}
  \caption{The expected plasma parameters for the proposed experiment at the NIF, LLNL. These are derived from experiments done at other  facilities \citep{tzeferacos2018laboratory}, rescaled to NIF laser drive conditions. The estimates for the Reynolds and magnetic Reynolds number follow \citep{Braginskii65} and \citep{Spitzer53}. However as discussed in \citep{Ryutov99} micro-instabilities may alter our estimates.}
  \label{Tab:Table1}
  \end{center}
\end{table}
We adopt these estimated parameters in order to assess the feasibility of a `cosmic ray acceleration platform' in the laboratory. The first step is the injection of cosmic ray particles, which can be implemented by replacing the CH foils used earlier~\citep{tzeferacos2018laboratory} by CD (deuterated plastic) foils. Within the turbulent plasma region, 3 MeV protons (with velocity $v_\textrm{p} \sim 8 \times 10^{-2}\,c$) will be  produced via D-D collisions in the counter-streaming plasma flows \citep{ProtonEmission}. Analogous to the astrophysical situation, these protons as they stream out from the plasma interact with the turbulent magnetic fields and should be accelerated by the 2nd-order Fermi process. We now model this interaction to predict the energy spectrum of the protons.

As each scatter changes the energy of a particle by a small fraction of its initial energy, this is a diffusion process in momentum space described by a Fokker-Planck transport equation for $f (p, t)$, the phase space density of the protons in the plasma 
\citep{EarlySolutionAttempt,TurbForStochAcc,ParticleAccelerationTheory,FermiToFokker}:
\begin{align}
\label{Eq:DiffusionEquation}
\left(\frac{\partial f}{\partial t}\right)_{\text{inner}}=&\frac{1}{p^2}\frac{\partial}{\partial p}\left(p^2 D_{p} \frac{\partial f}{\partial p}\right)
-\frac{f}{\tau_{\text{esc}}}\nonumber\\&+\frac{C_0\delta\left(p-p_0\right)\delta\left(t-t_0\right)}{4\pi p^2}.
\end{align}
Here $D_{p}$ is the momentum diffusion coefficient, $\tau_{\text{esc}}$ the escape time describing the loss of protons from the system and the last term describes instantaneous injection ($C_0$) of superthermal particles at time $t_0$ and momentum $p_0$. Note that spatial homogeneity is assumed in writing down Eq.~(\ref{Eq:DiffusionEquation}), however we will assume that the injection occurs only in the central region of the plasma. The particle distribution function $n$ is related to the phase space density $f$ through 
\begin{align}
\label{Eq:FtoN}
n(p,t)=4\pi p^2 f(p,t).
\end{align}

The diffusion equation~(\ref{Eq:FtoN}) has been solved for a variety of situations \citep{TurbForStochAcc,EarlySolutionAttempt,SecOrdFermSolAltern,Kaplan56,SolFermAcc} and we discuss below the appropriate solution for the proposed experiment.

The detectors used to measure the proton energy are located outside the plasma, hence the relevant distribution function to consider is that of the \emph{escaping} protons. This can be related to the particle distribution function of the protons inside the plasma by requiring that proton number be conserved after the injection, i.e. at $t>t_0$
\begin{align}
\label{Eq:OuterDiffusionEquation}
\left(\frac{\partial n}{\partial t}\right)_{\text{outer}}=\frac{n_{\text{inner}}}{\tau_{\text{esc}}},
\end{align}
where $f_{\text{inner}}$ is the solution of Eq.(\ref{Eq:DiffusionEquation}). 
As will be shown in \S \ref{sec:level3}, the escape time $\tau_{\text{esc}}$ is momentum-dependent so the phase space density inside and outside the plasma will be \emph{different}.

%----------------------------------------------------
\section{\label{sec:level3} The transport coefficients}
The Fokker-Planck transport coefficient for energy diffusion is \cite[e.g.][]{ParticleAccelerationTheory}
\begin{align}
\label{Eq:DefDiffCoeff}
D_{\epsilon}=\frac{\left<\left( \Delta \epsilon\right)^2\right>}{\Delta t},
\end{align}
where $\langle.\rangle$ denotes the average over scattering angles, and the energy change per scatter is given by the integral over the force exerted by the electric field fluctuations, i.e.
\begin{align}
\label{Eq:EnergyChange}
\langle\left(\Delta \epsilon\right)^2\rangle= \left<\left(e\int \mathbf{E}\mathbf{\cdot}\mathrm{d}\mathbf{s}\right)^2\right>\sim e^2\ell^2\langle\left(E_{||}\right)^2\rangle.
\end{align}
Here, %$e$ is the proton (electron) charge, 
$\mathbf{s}$ is the world-line of the particle and $\ell$ is the scale length of the turbulence cells in the plasma flow. These cells are defined by the scale on which the electric field, $\bf E$, is statistically de-correlated, close to the outer scale of the turbulent spectrum.

The appropriate Ohm's law reads:
\begin{align}
\label{Eq:GenOhmLawFinal}
\mathbf{E}=-\frac{\mathbf{u}\times\mathbf{B}}{c}-\frac{\mathbf{\nabla} P_{\rm e}}{n e} ,
\end{align}
where ${P_e}$ is the isotropic electron pressure, $n$ is the electron density and $\mathbf{u}$ is the electron velocity field. In principle Ohm's law contains possible additional contributions but these turn out to be small in the present case as e.g. the Reynolds and magnetic Reynolds numbers are large.
Note that the ratio of the two, the magnetic Prandtl number $P_{\rm m}=R_{\rm m}/Re$ is much larger than unity at the NIF (see Table I), confirming that the plasma is in an astrophysically relevant regime.

For the assumed conditions at the NIF, the pressure term is insignificant, being of ${\cal O}(10^{-2})$.  We will retain it nevertheless as it is relevant for conditions achievable at other facilities such as OMEGA \citep{soures}.

Returning to the Fokker-Planck transport coefficient~(\ref{Eq:DefDiffCoeff}), we find by combining Eqs.(\ref{Eq:EnergyChange}) and (\ref{Eq:GenOhmLawFinal})
\begin{align}
\label{Eq:EnergyChangeStart}
\left<\left(\Delta \epsilon\right)^2\right>=&e^2\ell^2\left<\left(\frac{u^2B^2}{c^2}\sin^2(\theta)\right)\right>+e^2\ell^2\left<\left(\frac{\mathbf{\nabla} P_{\rm e}}{n e}\right)^2\right>\nonumber\\
&+2e^2\ell^2\left<\left(\frac{uB}{c}\left|\frac{\mathbf{\nabla} P_{\rm e}}{n e}\right|\sin(\theta)\cos(\phi)\right)\right>,
\end{align}
with $\theta$ the angle between $\mathbf{u}$ and $\mathbf{B}$ and $\phi$ the angle for the inner product of the two terms in (\ref{Eq:GenOhmLawFinal}) after squaring.

Treating the electrons as an ideal gas, one finds for the electron pressure term: $\mathbf{\nabla} P_{\rm e}= e \mathbf{\nabla} (nT)$. Due to the large electron thermal conduction, the temperature remains approximately constant across the plasma \citep{tzeferacos2018laboratory}, hence pressure fluctuations are due only to density variations.
This also implies that the angles $\theta$ and $\phi$
are uncorrelated, hence the averaging gives
\begin{align}
\label{Eq:AveragedECS}
\left<\left(\Delta \epsilon\right)^2\right>=\ell^2\left[\frac{4}{3}\frac{e^2}{c^2}u^2B^2+e^2T^2\left(\frac{\mathbf{\nabla} n}{n}\right)^2\right].
\end{align}
The relevant time scale for this energy change depends on the parameters of the system. This can be in one of two  regimes --- ballistic escape or true diffusion --- according to how the pitch angle scattering time compares to the time a proton needs to escape the plasma.

The relevant time scale determining the Fokker-Planck coefficient (\ref{Eq:DefDiffCoeff}) is the time it takes a proton to cross a turbulent cell, taken to be of the order of the grid size i.e., $\Delta t \sim \ell/v_\mathrm{p}$. For the magnetic field values expected in the NIF experiment (Table I), the proton gyro-radius is $\rg\sim 0.2$ cm, i.e. larger than the turbulent cell scale, so particles are \emph{unmagnetised}. Hence proton propagation can be described as a random walk until escape  from the plasma. Since the protons are non-relativistic, we have
\begin{align}
\label{Eq:TimeScale}
\Delta t \sim \frac{\ell}{v_\mathrm{p}}=\frac{m_\mathrm{p} \ell}{p},
\end{align}
where $p=m_\mathrm{p} v_\mathrm{p}$ is the proton momentum.

Moreover in this case,
$(\Delta \epsilon)^2 = (p^2/m_\mathrm{p}^2) (\Delta p)^2$, so the momentum diffusion coefficient is from Eqs.(\ref{Eq:AveragedECS}) and (\ref{Eq:TimeScale}):
\begin{align}
\label{Eq:DiffusionCoefficient}
D_p=
\frac{\left<\left(\Delta p\right)^2\right>}{\Delta t}\sim\
\frac{\ell}{c}\left[\frac{4e^2B^2}{3}\frac{u^2}{c^2}+e^2 T^2\left(\frac{\mathbf{\nabla} n}{n}\right)^2\right]\frac{m_\mathrm{p} c}{p}.
\end{align}
We note the dependence $D_p\propto p^{-1}$.
%of the momentum diffusion coefficient.
In the traditional moving cloud picture, diffusion is \emph{less} efficient for particles with higher momentum because the difference in probability for head-on and over-taking collisions is then smaller. In the present setup with MHD turbulence, particle trajectories are simply aligned with the electric field for a shorter time, provided they remain non-relativistic. 

%As the collision rates are functions of the relative velocity, hence nearly equal for particles that are much faster than the time it takes to cross a turbulence cell. The diffusion coefficient falls off at large $p$ because such particles do not fulfill the resonance condition for scattering, hence diffusion becomes inefficient. For true diffusion, the protons will not simply stream out of the plasma as in the present case, however the time scales estimated above are still relevant in our framework.

Next we need the spatial diffusion coefficient in order to determine the diffusive escape-time scale. The mean free path is the distance a proton travels before it is deflected by an angle $\pi/2$ in a time $\tau_{90}$ i.e. $\lambda = v_\mathrm{p}\tau_{90}$. This can be estimated by treating the change in angle by each turbulence cell as a random walk process so the time to be scattered $\pi/2$ away from the initial direction is,
\begin{align}
\label{Eq:T90Time}
\tau_{90}\sim \frac{\rg}{\wg \ell}= \frac{m_\mathrm{p}^2c^3}{\ell e^2 B^2}\frac{p}{m_\mathrm{p} c},
\end{align}
where $\wg = v_{\rm p}/\rg = eB/mc$, with $B$ given by its rms value.
The spatial diffusion coefficient is thus:
\begin{align}
\label{Eq:SpatDiffCoeff}
D_x=\frac{\left<\left(\Delta x\right)^2\right>}{\Delta t}\sim\frac{1}{3}\frac{\lambda^2}{\tau_{90}}=\frac{m_\mathrm{p}^2 c^5}{3e^2 \ell B^2}\left(\frac{p}{m_\mathrm{p} c}\right)^3.
\end{align}

The spatial and momentum diffusion coefficients are related through:
\begin{align}
\label{Eq:DPDXconnection}
D_p D_x=\frac{m_\mathrm{p}^2c^4}{3e^2B^2}\left[\frac{4e^2B^2}{3}\frac{u^2}{c^2}+e^2 T^2\left(\frac{\mathbf{\nabla} n}{n}\right)^2\right]\left(\frac{p}{m_{\rm p}c}\right)^2.
\end{align}

The escape time $\tau_{\text{esc}}$ is the time it takes a particle to diffuse out of the turbulent region of size $L$: 
\begin{align}
\label{Eq:EscapeTime}
\tau_{\text{esc}}=\frac{L^2}{D_x}=\frac{3e^2}{m_{\rm p}^2c^5}\ell\,L^2B^2\left(\frac{p}{m_{\rm p} c}\right)^{-3}.
\end{align}
This shows that particles with higher momentum get lost more efficiently, as is expected, since such particles stream out of the plasma faster, in addition to having a longer mean free path.
Consequently, the mean momentum of particles outside the plasma will be higher due to the biased escape of mostly fast particles. Moreover, slower particles remain inside the plasma longer, accounting for their stronger acceleration.

As noted before, the system can be in two different limits: true diffusion or ballistic escape. The regimes are characterised by comparing the intrinsic time-scales of the system. Using the parameters from Table \ref{Tab:Table1} the angular scattering time as defined by Eq.(\ref{Eq:T90Time}) is
\begin{align}
\label{Eq:AngularScattTime}
\tau_{90}\sim \frac{\rg}{\wg \ell}\sim 1.5 \times 10^{-10}\,\text{s} \left(\frac{B}{1.2\text{ MG}}\right)^{-2} \left(\frac{\ell}{0.1\text{ cm}}\right)^{-1},
\end{align}
while the time needed to diffuse out of the plasma is:
\begin{align}
\label{Eq:EscapeTimeEstimate}
\tau_{\text{esc}}=\frac{L^2}{D_x}\sim 5.5\times10^{-10}\,\text{s} \left(\frac{B}{1.2\text{ MG}}\right)^{2}\left(\frac{\ell}{0.1\text{ cm}}\right).
\end{align}
These must be compared with the time a proton would take to cross the plasma in the absence of magnetic fields:
\begin{align}
\label{Eq:CrossingTime}
t_{\text{cross}}=\frac{L}{v_{\rm p}}=1.7\times10^{-10}\,\text{s}.
\end{align}
All the time scales above are of the same order, indicating that the conditions at NIF will be close to the transition from ballistic escape to the diffusive regime. We can optimistically also consider the case when $B$ is larger by a factor of $\sim3$ (chosen to guarantee a factor of $O(100)$ change in the time-scales). Then $B\sim 3.6$~MG and
\begin{align}
\label{Eq:DiffusiveComparisonTimes}
\tau_{90}\sim 3.4\times 10^{-11}\,\text{s} < \tau_{\text{esc}}=2.5\times10^{-9}\,\text{s}.
\end{align}
We may now safely assume true diffusion rather than ballistic escape.

%-------------------------------------------------
\section{\label{sec:level4} Solving the diffusion equation}
Given the diffusion and loss coefficients that we have derived, we can employ an analytical solution~\citep{SolFermAcc} obtained under the following assumptions. 
First, the initial and final proton distributions are assumed to be isotropic so the distribution function depends only on $p=|\mathbf{p}|$. In the present case, even if the turbulent plasma is produced by the collision of two counter propagating flows, their centre-of-mass is at rest, which suggests that the D-D proton emission is isotropic on average. However each individual D-D pair does not necessarily have a stationary centre-of-mass due to the temperature of the plasma jets so the initial energy distribution is not mono-energetic. We will discuss the implication of this in the next section. Experimental data show that the properties of turbulence are uniform within the interaction region of size $L$ \citep{tzeferacos2018laboratory} so any spatial dependence may safely be neglected. Second, the plasma must be magnetised, which is the case for electrons in the proposed experiment (however, the ions are only weakly magnetised). Third, the proton energies must be relativistic. Although this is \emph{not} the case here, the analytical solution~\citep{SolFermAcc} holds as long as $D_p D_x \propto p^2$ and as shown earlier this relation remains valid even in the non-relativistic case.

Noting that $D_p$ and $\tau_{\text{esc}}$ are \emph{constant} in time, the solution of Eq.(\ref{Eq:DiffusionEquation}),
taking $C_0=1$, is then \citep{Kaplan56,SecOrdFermSolAltern,SolFermAccOld,SolFermAcc}:
\begin{align}
\label{Eq:Solution}
n_{\text{inner}}=\frac{2\hat{p}^2\sqrt{\Psi}}{\sqrt{k\tau}\left(1-\Psi\right)}e^{-\frac{\left(\hat{p}^3+\hat{p}_0^3\right)\left(1+\Psi\right)}{3\sqrt{k\tau}\left(1-\Psi\right)}}\text{I}_0\left[\frac{4\left(\hat{p}
\hat{p}_0\right)^{\frac{3}{2}}\sqrt{\Psi}}{3\sqrt{k\tau}\left(1-\Psi\right)}\right],
\end{align}
where $\hat{p}$ and $\hat{p}_0$ are dimensionless momenta, e.g., $\hat{p}=p/m_{\rm p} c$, and $\text{I}_0$ is the modified Bessel function of the first kind. The function $\Psi$ is defined as
\begin{align}
\label{Eq:Psi}
\Psi(t,t_0)=\exp\left[-6\sqrt{\frac{k}{\tau}}\left(t-t_0\right)\right],
\end{align}
with
\begin{align}
\label{Eq:KDefinition}
k=\frac{D_p}{m_{\rm p}^2 c^2}\frac{p}{m_{\rm p} c},
\end{align}
and,
\begin{align}
\label{Eq:DefTau}
\tau=\tau_{\text{esc}}\left(\frac{p}{m_{\rm p} c}\right)^3.
\end{align}

\begin{figure*}
\centering
\begin{subfigure}{.5\linewidth}
  \centering
  \includegraphics[width=.95\linewidth]{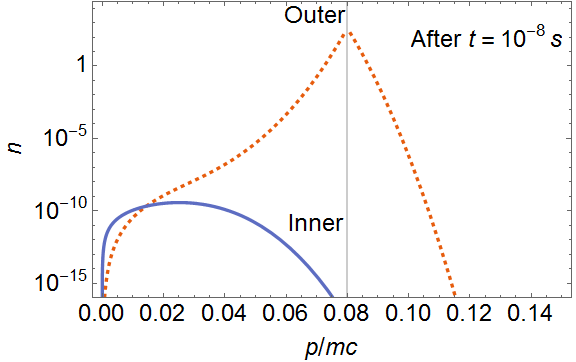}
  \caption{}%After $t=10^{-8}$ $s$.
  \label{fig:sub1}
\end{subfigure}%
\begin{subfigure}{.5\linewidth}
  \centering
  \includegraphics[width=.95\linewidth]{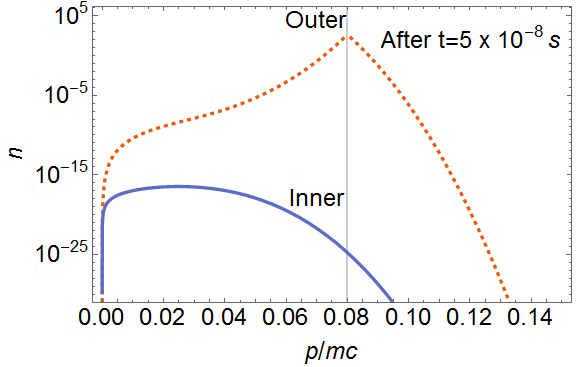}
  \caption{}%After $t=5\times 10^{-8}$ $s$.
  \label{fig:sub2}
\end{subfigure}
\\
\begin{subfigure}{.5\linewidth}
  \centering
  \includegraphics[width=.95\linewidth]{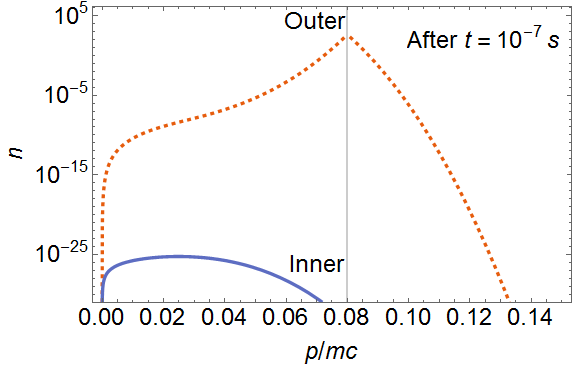}
  \caption{}%After $t=10^{-7}$ $s$.
  \label{fig:sub3}
\end{subfigure}%
\begin{subfigure}{.5\linewidth}
  \centering
  \includegraphics[width=.95\linewidth]{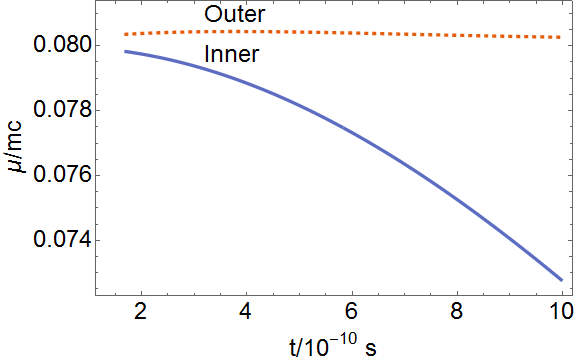}
  \caption{}%Time dependence of the mean momentum.
  \label{fig:sub4}
\end{subfigure}
\caption{The particle distribution functions inside (solid) and outside (dashed) the plasma. The vertical line indicates the initial proton momentum. The outer distribution does not change significantly after $\sim 10^{-8}$~s (c.f. panels \ref{fig:sub1}-\ref{fig:sub3}) while the inner distribution falls off quickly. Panel~\ref{fig:sub4} shows the time dependence of the mean momentum.}
\label{Fig:InnerOuter}
\end{figure*}

In order to now determine the outer distribution function, $n_{\text{outer}}$, we can simply integrate Eq.(\ref{Eq:OuterDiffusionEquation}) to find:
\begin{align}
\label{Eq:OuterDistrSolution}
n_{\text{outer}}(p,p_0,t,t_0)=\int_{0}^{t} \frac{n_{\text{inner}}(p,p_0,t',t_0)}{\tau_{\text{esc}}}\,\text{d}t'.
\end{align}
Both distributions are shown in Fig.~\ref{Fig:InnerOuter}. Due to the scaling $\tau_{\text{esc}}\propto p^{-3}$, escape from the plasma is biased towards higher momentum protons. This explains the decreasing mean momentum inside the plasma, since only the slower particles remain after a time comparable to $\tau_{\text{esc}}$. For the same reason the mean momentum outside the plasma is higher than on the inside.
The momentum spectra shown in Fig.~\ref{Fig:InnerOuter} were obtained by substituting the values for the plasma conditions given in Table~\ref{Tab:Table1}.

Momentum diffusion also changes the mean momentum of the proton distribution. We can obtain analytically the mean proton momentum \emph{inside} the plasma:
\begin{align}
\label{Eq:MeanValue}
\mu_{\text{inner}} = & \frac{3^{1/3}}{\sqrt{k\tau}}\left[k^2\tau^2\frac{\left(1-\Psi\right)^4}{\left(1+\Psi\right)^4}\right]^{1/3}\Gamma\left(\frac{4}{3}\right) \times \nonumber\\
	&\exp\left[{-\frac{4\Psi \hat{p}_0^3}{3\sqrt{k\tau}\left(1-\Psi^2\right)}}\right]\text{L}_{-\frac{4}{3}}\left[\frac{4\Psi \hat{p}_0^3}{3\sqrt{k\tau}\left(1-\Psi^2\right)}\right],
\end{align}
where $\text{L}_{-\frac{4}{3}}$ are Laguerre polynomials, and $\Gamma$ is the Gamma function. However the mean momentum outside can only be calculated numerically.

%-----------------------------------------
\section{\label{sec:level5} Experimental feasibility}
The two relevant quantities that can be measured in a possible experiment are the shift in the mean energy and e.g. the full width at half maximum (FWHM) of the proton distribution. Of particular interest is the outer distribution, since that is where the detector is located. Given the available diagnostics, the measurement is essentially time-integrated, being the integral of Eq.~(\ref{Eq:OuterDistrSolution}) from the initial time to infinity.

In practice, it is sufficient to integrate up to a time late enough such that a significant portion of the protons have escaped from the plasma, indicated by the inner distribution dropping to near zero. A time of order $10^{3}\tau_\textrm{esc}$ proves sufficient as will be shown later. Also, the lower bound must be modified, due to the delta function nature of the distribution at $t = t_0$. Taking into account that the shortest time-scale on which a particle can exit the plasma is just crossing time, the lower bound on the integral can be chosen to be $t_0 \sim \tau_{\text{esc}}$. Then we obtain for the mean momentum of the escaping protons:
\begin{align}
\label{Eq:FinalMeanMomentum}
\mu_\textrm{outer}\sim \int_0^1 \hat{p}\int_{10^{-13}}^{10^{-7}} \frac{n_{\text{inner}}}{\tau_{\text{esc}}} \text{d}t\,\text{d}\hat{p} =0.08\, c,
\end{align}
which corresponds to a mean energy of 3.01~MeV.
As expected, this is higher than at injection --- by 10~keV which is $\sim 3\%$ of the initial proton energy. The upper limit $\hat{p}=1$ was chosen to ensure  that $\hat{p}\gg \hat{p}_0$ and thus out of reach of the acceleration mechanism. This follows from the Hillas criterion \citep{HillasLimit}, which provides an upper limit on the maximum energy gain by comparing the system size with the particle gyro-radius. The Hillas limit in our case is
\begin{align}
E_{\text{Hillas}}=eBL\frac{u}{c}=0.24\,\text{MeV},
\end{align}
which is larger than the predicted energy gain of $\sim 0.01$~MeV. The expected width of the distribution is $\Delta E_{\text{FWHM}} \sim0.4\,\text{MeV}$, i.e. $\sim 15\%$ of the proton energy.

Applying the same approximations to the inner distribution, we find for the mean energy $\mu_\textrm{inner}\sim 0.6\,\text{MeV}$. This distribution, however, is not relevant any more, as the overall probability to find a particle inside the plasma after such long times has dropped to
\begin{align}
\label{Eq:OverallParticleNumberInside}
N_{\text{inner}}=\int_{0}^{1}n_{\text{inner}}\,\text{d}\hat{p}\sim 10^{-28}.
\end{align}
Here the upper limit was chosen such that it is much bigger than the mean initial momentum $p_0$ and thus out of reach of the acceleration process. 

Note that the numbers above correspond to impulsive injection of one particle ($C_0=1$) at one point in time. Our result can be simply extended for multiple impulsive injections by appropriate superposition of the solution. If the time during which particles are injected is shorter than both the plasma and detector accumulation time, then the resultant proton spectrum is just that for a single impulsive injection scaled by a multiplicative factor.

\begin{figure*}
\centering
\begin{subfigure}{.5\linewidth}
  \centering
  \includegraphics[width=.95\linewidth]{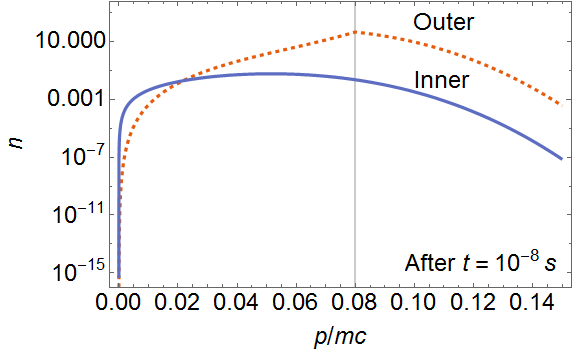}
  \caption{}%After $t=10^{-8}$~s.
  \label{fig:sub21}
\end{subfigure}%
\begin{subfigure}{.5\linewidth}
  \centering
  \includegraphics[width=.95\linewidth]{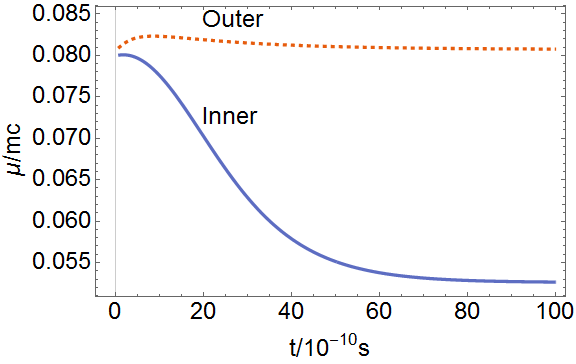}
  \caption{}%Mean momentum evolution.
  \label{fig:sub22}
\end{subfigure}
\caption{The particle distribution functions inside (solid) and outside (dashed) the plasma taking $B=3.6$ MG. The vertical line indicates the initial proton momentum. Panel \ref{fig:sub22} shows the time dependence of the mean momentum.}
%beginning at the crossing time.}
\label{Fig:InnerOuter2}
\end{figure*}

We do the same analysis assuming a  higher peak magnetic field of $B \sim 3.6 \text{ MG}$, which as noted before guarantees the system to be in the diffusive regime. As Fig.~\ref{Fig:InnerOuter2} shows, both the inner and outer distributions start off at the initial momentum for times around the escape time of these particles. Subsequently the mean momentum  of the outer distribution increases due to the biased escape, and decreases only when the time is long enough for slower particles to escape the plasma. 
%The inner distribution decays continuously. 
%The behaviour is similar to before although the interesting statistical measures are rather different.

As expected, the mean proton energy is now higher than before: $\Delta \mu_{\epsilon}\sim 200\,\text{keV}$.
%That was to be expected because of a combination of stronger fields and longer escape times. The mean energy change is of the order of
%\begin{align}
%\label{Eq:EnergyCHangeHigherB}
%\Delta \mu_{\epsilon}\sim 200\,\text{keV}.
%\end{align}
The same is true for the FWHM, which increases to: $\Delta E_{\text{FWHM}}\sim 1.2\,\text{MeV}$.
This is consistent with the Hillas limit, which for the changed plasma parameters reads $E_{\text{Hillas}}=0.71\,\text{MeV}$.

Finally, as mentioned earlier, we need to  consider that the protons are not all injected with the same energy. However the thermal broadening is expected to be 30-40~keV \citep{ThermalSpread} (see also \cite{ThermalSpreadold}), i.e. an order of magnitude smaller than our calculated effect. On top of the thermal broadening there is a second effect related to the turbulent motion which can be estimated by noting that the relative change in energy is $\Delta E/E=4u/v_p \sim 0.1$, i.e. $\Delta E \sim 300\,\text{keV}$ which is smaller than the expected broadening due to stochastic acceleration.

In ascribing any measured energy shift and spectral broadening to the 2nd-order Fermi mechanism, it must be noted that target charging, generation of static electric fields due to the escape of hot electrons and energy loss by collisions can all obscure the signal \citep{PhysRevLett.114.215002,doi:10.1063/1.1320467}. Other experiments with the proposed set-up are evaluating these effects and indicate that the resulting distortions are in fact negligible \citep{ChenPreprint}.

We can interpret the spread due to turbulent motion of $\Delta E=300keV$, as a lower bound for detection. Therefore given plasma conditions as in Table I, and certainly for a 3 times stronger magnetic field, 2nd-order Fermi acceleration of D-D fusion protons should be measurable at the NIF.

To summarise, we have presented a suitable experimental set-up for measuring stochastic acceleration of protons in a turbulent plasma. If realised, this would provide a platform where basic physical processes related to the classic Fermi theory of cosmic ray acceleration can be directly tested and validated against numerical simulations.
We have demonstrated that the unique experimental capabilities available at NIF offer a potential route to explore collisionless magnetised transport, where unlike in previous studies \citep{ChenPreprint} the crossing time exceeds the scattering/isotropisation time. For the experimental conditions we consider, the transition occurs at turbulent field strengths of $B \lesssim 3$ MG, which are theoretically achievable using the experimental set-up of \cite{tzeferacos2018laboratory}. This regime may have practical implications for the acceleration of cosmic rays in the presence of sub-Larmor scale turbulent fields in astrophysical systems \cite[e.g.][]{Bell2004}.

\section*{Acknowledgements:}
We would like to thank John Foster for valuable input. The research leading to these results has received funding from AWE plc. and the Engineering and Physical Sciences Research Council (grant numbers EP/M022331/1, EP/N014472/1 and EP/P010059/1). SS acknowledges a Niels Bohr Professorship awarded by the Danish National Research Foundation. \copyright British Crown Copyright 2018/AWE

% susie put cite commands here, don't bother with citet etc just yet.

%\bibliographystyle{jpp}

\bibliography{main}

\end{document}